\documentclass{optica-article}

\journal{opticajournal} 

\articletype{Research Article}
\usepackage{graphicx,epsf,amsmath}
\usepackage{dcolumn}
\usepackage{bm,bbm}
\usepackage{enumerate}
\usepackage[normalem]{ulem}
\usepackage{upgreek}
\usepackage{braket}
\usepackage{siunitx}

\usepackage{float}
\usepackage{dsfont}
\usepackage{mathtools}

\usepackage{graphicx}
\usepackage{hyperref}
\usepackage{cleveref}
\usepackage{subcaption} 
\usepackage{lineno}

\begin{document}
\title{Demonstration of a reconfigurable quantum network architecture suitable for ground-to-space communication}

\author{St\'ephane Vinet,\authormark{1,*} Duncan England,\authormark{2} Chang-qing Xu,\authormark{3} and Thomas Jennewein\authormark{1,4}}

\address{\authormark{1}Institute for Quantum Computing and Department of Physics \& Astronomy,
University of Waterloo, 200 University Ave W, Waterloo, Ontario, N2L 3G1, Canada\\
\authormark{2}National Research Council of Canada, 100 Sussex Drive, Ottawa, ON, K1A 0R6, Canada\\
\authormark{3}Department of Engineering Physics, McMaster University,1280 Main Street West, Hamilton, ON, L8S 4L8, Canada\\

\authormark{4}Department of Physics, Simon Fraser University, 8888 University Dr W, Burnaby, BC, V5A 1S6, Canada}
\email{\authormark{*}svinet@uwaterloo.ca}

\begin{abstract*} 
  We experimentally demonstrate a reconfigurable quantum network architecture suitable for integrating satellite links in metropolitan quantum networks. The network architecture is designed such that once a satellite is in range, it is configured in a multipoint-to-point topology where all ground nodes establish entanglement with the satellite receiver using time multiplexing to optimize long-distance transmission. Otherwise, the satellite up-link can be rerouted to the ground nodes to form a pair-wise ground network. Leveraging both the time and frequency correlations of our photon-pair source, we demonstrate an increased coincidence-to-accidental ratio without additional resource overhead in a five-node network. To contextualize these experimental findings, we project their performance in a quantum key distribution scenario and outline a feasible route towards field deployment, using integrated photonics to enable network integration of up to 72 users.
\end{abstract*}
\section{Introduction}
The development of the quantum internet will enable advanced applications in quantum cryptography, quantum-enhanced metrology, and quantum computing \cite{Broadbent,Nickerson}. A fundamental building block of this vision is the ability to share quantum entanglement between distant nodes \cite{KhodadadKashi_2025}. To-date, quantum networks have mostly been restricted to metropolitan distances with few predominantly static nodes, constraining their scalability and adaptability for broader deployment \cite{wengerowskyentanglementbased2018,Joshi,darpa, Peev_2009, chen_implementation_2021}. These limitations stem largely from losses in optical fibres and the significant infrastructure required to deploy and maintain terrestrial links over long distances. Satellites provide a promising platform to circumvent these challenges and distribute entanglement over global distances. However, there are significant challenges to integrate them seamlessly with terrestrial networks \cite{2017Natur.549...43L,PhysRevLett.119.200501, 2018Liao}. In particular, satellite links experience significant attenuation and intermittent connectivity during orbital passes \cite{PRXQuantum.5.030101, Liorni_2019}. To overcome such limitations, some of the authors have proposed in \cite{vinet_reconfigurable_2025} a reconfigurable network architecture.
The network dynamically reconfigures its topology based on the satellite's accessibility to optimize the limited availability of the satellite. During a satellite pass, the network operates in a multipoint-to-point configuration, enabling all ground nodes to establish entanglement with the satellite simultaneously. When the satellite is unavailable, the satellite up-link is rerouted via an optical switch to the ground nodes, transforming the network into a pair-wise connected ground topology. This reconfiguration ensures efficient and continuous operation across both configurations with minimal modifications to the infrastructure. Furthermore, the performance of each configuration is improved using time and frequency multiplexing. 
\par In this paper, we experimentally demonstrate this network reconfiguration using correlated photon pairs generated from a periodically poled magnesium-oxide doped lithium niobate (PPLN) waveguide specifically designed for the Quantum Encryption and Science Satellite (QEYSSat) up-link scenario \cite{10.1117/12.2041693}. Broadband non-degenerate photon pairs at 1550 and 785 nm are generated from spontaneous parametric down-conversion (SPDC) to optimize both the fibre and free-space link. Using time multiplexing, we demonstrate a linear performance increase for the up-link configuration with minimal satellite hardware requirements. In particular, a pulsed pump paired with a frequency-to-time mapping (FTM) assigns a distinct time delay to each spectral channel enabling demultiplexing on a single detector and thus reducing the satellite payload requirements \cite{Davis:17,KhodadadKashi_2025}. This resource alleviation extends to the ground configuration where only one detector per user is required. Indeed, the pulsed pump allows for channel discrimination by defining fixed time slots for the arrival of each channel at the detector thus reducing the number of accidental coincidences \cite{wengerowskyentanglementbased2018}.
\section{Experimental setup}\label{sec:exp}
\begin{figure}
    \centering
    \includegraphics[width=\textwidth]{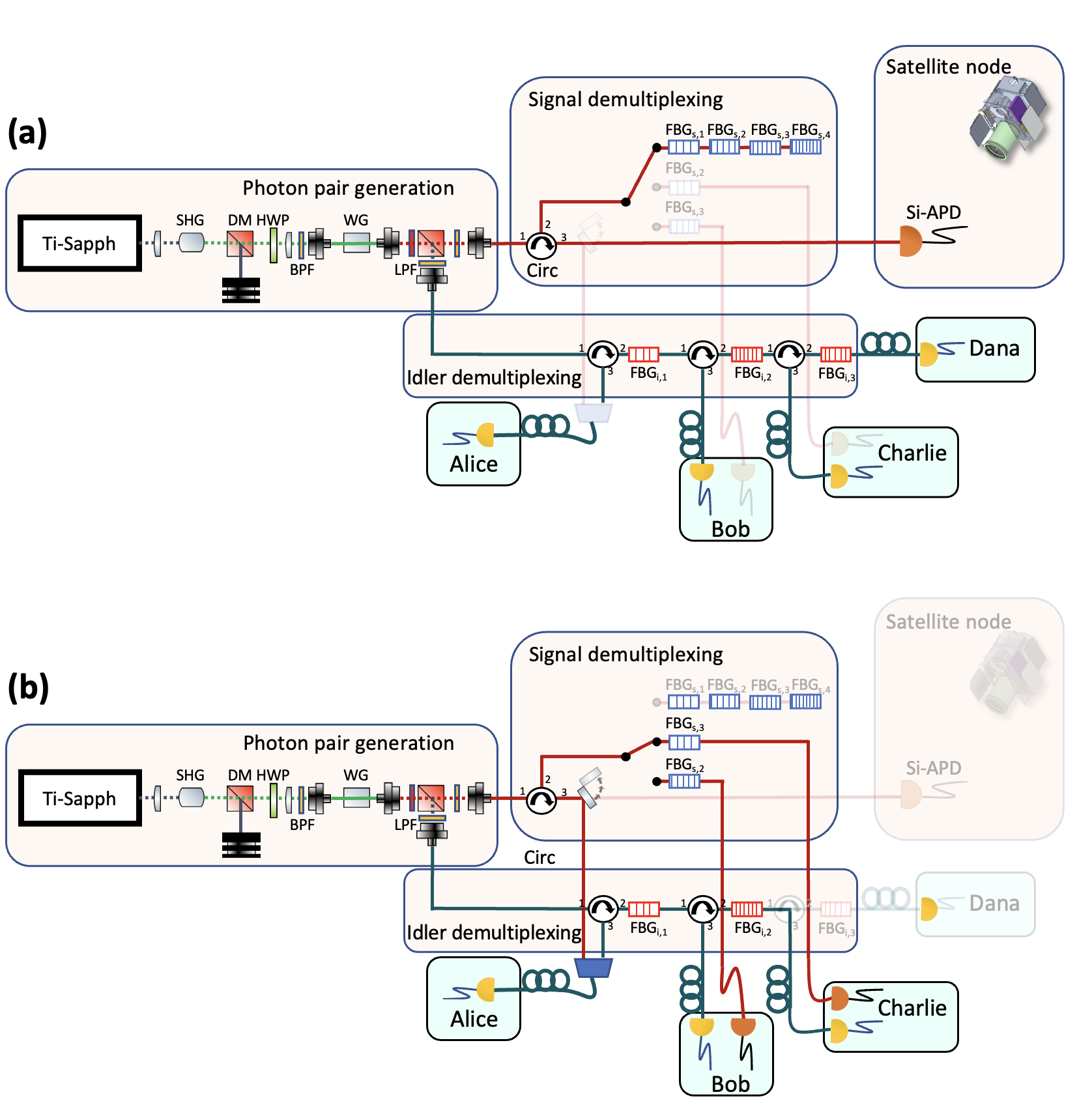}
    \caption{Experimental implementation of the reconfigurable quantum network for \textbf{(a)} the satellite configuration and \textbf{(b)} the ground configuration. To emphasize the differences between the two configurations, optical elements not used in each configuration are depicted in grey. (SHG: second-harmonic generation, DM: dichroic mirror, HWP: Half-wave plate, BPF: band-pass filter, WG: waveguide, LPF: long-pass filter, Circ: circulator, FBG: fibre Bragg grating, SNSPD:  superconducting nanowire single photon detector, Si-APD: silicon avalanche photodetector) }
    \label{fig:setup}
\end{figure}

The experimental setup is shown in Fig.~\ref{fig:setup}. A mode-locked titanium-sapphire laser (\textit{Tsunami Spectra-Physics}) with a repetition rate of 80 MHz produces 1042 nm light, which is converted in a Barium borate (BBO) crystal (\textit{Castech Inc.}) via second-harmonic generation and filtered down to a $\sim 2$ nm bandwidth centered around 521.4 nm. These upconverted photons then pump a 15 mm long PPLN waveguide (\textit{HC Photonics Corp.}) such that broadband photon pairs at 787.5 nm (signal) and 1543.2 nm (idler) are produced via non-degenerate type-0 SPDC. PPLN waveguides offer high brightness, compactness, and compatibility with integrated devices \cite{2020Kuo}. After correction for losses, the brightness of the SPDC source was estimated to be $>7.8\times 10^7$ pairs/sec/mW.
A temperature controller was used to maintain the PPLN’s temperature at 30 $^\circ$C such that the phase matching of the SPDC process is optimized to maximize photon generation within the QEYSSat passband chosen for high atmospheric transmission and compatibility with high efficiency space-ready detectors \cite{Brambila:23}. Signal and idler photons are separated at a dichroic mirror and coupled into polarization-maintaining (PM) fibres. The signal fibre output is connected to two different detection scenarios depending on the satellite's availability. For a satellite up-link, the idler photons remain on the ground and are routed, in both detection scenarios, to the various users via wavelength division multiplexing (WDM) using concatenated telecom FBGs (\textit{O/E Land Inc.}) and optical circulators (\textit{OZ Optics}). 

\par
\begin{figure}
    \centering
    \includegraphics[width=\linewidth]{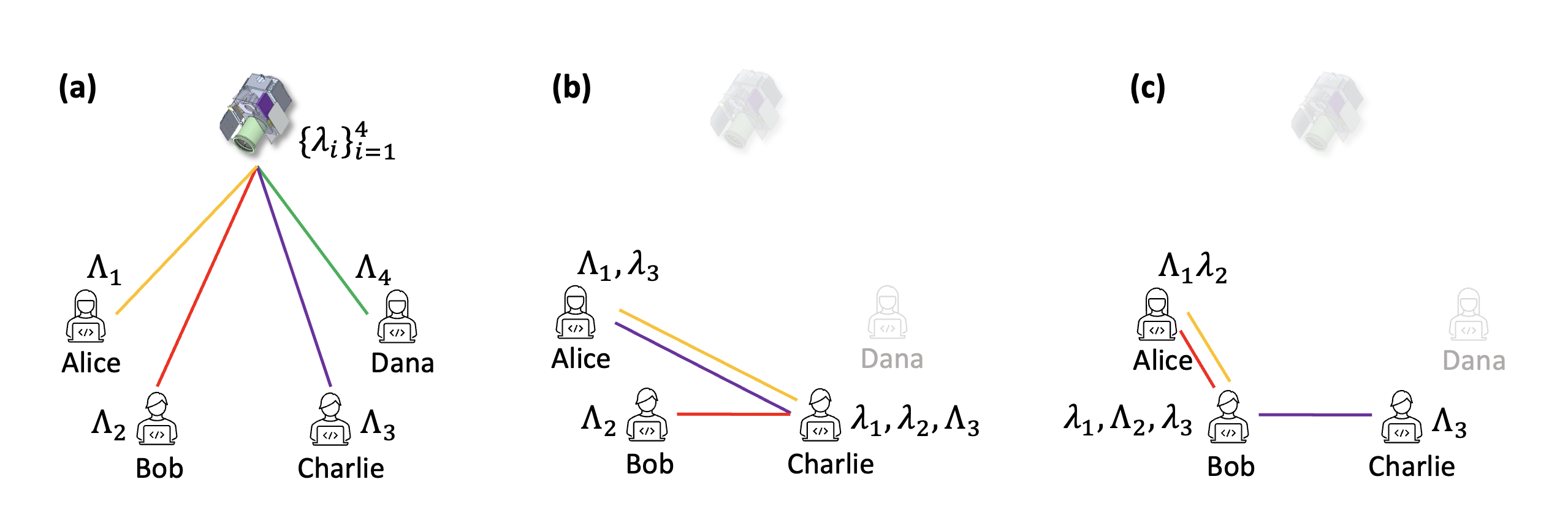}
    \caption{The quantum‐correlation layer for each switch position (top, middle and bottom) in the signal demultiplexing stage. \textbf{(a)} corresponds to the satellite configuration, while \textbf{(b)} and \textbf{(c)} correspond to the ground configuration.  }
    \label{fig:correlation}
\end{figure}

 For the satellite configuration shown in Fig.~\ref{fig:setup}a, a FTM unit enables a time-resolved detection of the distinct frequency channels on the same Si-APD detector. This FTM is implemented using a fibre Bragg grating (FBG) array that inscribes a distinct time of arrival to each spectral channel, enabling multipoint-to-point communication with the satellite as depicted in Fig.~\ref{fig:correlation}a. For the ground network shown in Fig.~\ref{fig:setup}b, a fully-connected topology between Alice, Bob and Charlie is established by wavelength division multiplexing (WDM) using individual FBGs, while Dana is excluded due to experimental resource constraints. Although the connections in Fig.~\ref{fig:correlation}b-c are established asynchronously using an active switch, they could be implemented simultaneously by adding an optical circulator to passively demultiplex all three signal wavelengths. The wavelength allocation for each network configuration is provided in Table~\ref{tab:alloc}. To optimize detection efficiencies around both $780$ and $1550$ nm, Bob and Charlie were equipped with a Si-APD (\textit{Excelitas}) and SNSPD detector (\textit{Quantum Opus}). This requirement could be overcome by using a broadband detector \cite{Gourgues:19, 10.1063/5.0046057,doi:10.1021/acsphotonics.0c00433}, thus eliminating the need for multiple detectors tailored to specific wavelengths, thereby reducing the system complexity. To demonstrate this scalable approach, the $1550$ and $780$ nm photons, corresponding respectively to channels $\Lambda_1$ and $\lambda_2 $ in Fig.~\ref{fig:correlation}b (and similarly $\Lambda_1$ and $\lambda_3 $ in Fig.~\ref{fig:correlation}c), are combined into the same single-mode fibre (\textit{Thorlabs SMF-28}) using a 1550/780 nm wavelength division multiplexer (\textit{Seagnol Photonics}) and jointly recorded on Alice's SNSPD detector.
Although the 780 nm photons are slightly multimode and suffer significant optical loss in the telecom fibre, Meyers-Scott \textit{et al.} \cite{Meyers} demonstrated the high-fidelity transmission is nonetheless possible and Vinet \textit{et al.} \cite{vinet_reconfigurable_2025} showed that positive nonzero key rates can be generated for up to 32 km, enough to span a metropolitan area network. 

\section{Experimental results}\label{sec:results}
\begin{figure}
\begin{subfigure}{0.49\textwidth}
\captionsetup{font=bf}
    \centering
    \includegraphics[width=\textwidth]{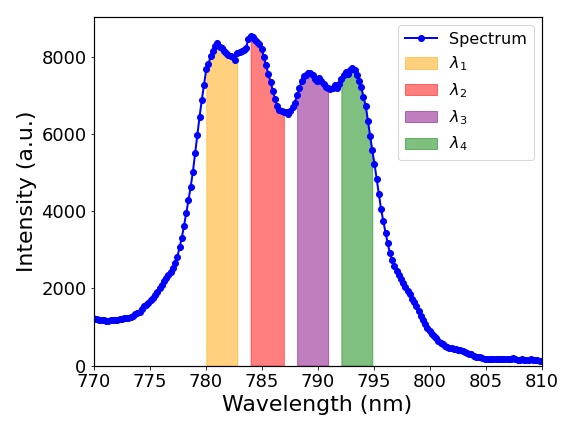}
    \caption{}
    \label{fig:signal}
    \end{subfigure}
    \begin{subfigure}{0.49\textwidth}
    \captionsetup{font=bf}
    \includegraphics[width=\textwidth]{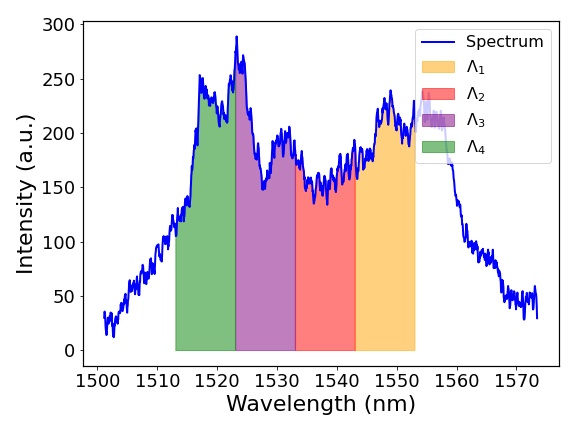}
 \caption{}
\end{subfigure}
\caption{The measured spectrum of the signal \textbf{(a)} and idler \textbf{(b)} photons. Wavelength correlated channels $\{\lambda_{i},\Lambda_{i}\}$ are denoted with the same color in both spectrums and reflected by $FBG_{s(i)}$ respectively in Fig.~\ref{fig:setup}. 1 nm guard bands are used between the signal channels $\{\lambda_i\}$ to minimize cross-talk between the channels.}
\label{fig:spectrum}

\end{figure}

To demonstrate the multiplexing advantage, we measured the coincidence-to-accidentals ratio (CAR) in both scenarios using up to four correlated channels specified within the broadband spectrum of the SPDC source, see Fig.~\ref{fig:spectrum}. 
\begin{table}[h]
    \centering
    \captionsetup{justification=centering}
    \caption{Wavelength allocation: wavelength correlated pairs are denoted as $\{\lambda_{i},\Lambda_{i}\}$ for signal and idler respectively. The first column corresponds to the position of the switch in the signal demultiplexing stage as identified in Fig.~\ref{fig:setup}.}
    \begin{tabular}{|c|c|c|c|c|c|}
    \hline
      Switch &Network  & \multicolumn{2}{c|}{Satellite configuration}& \multicolumn{2}{c|}{Ground configuration}\\
    position &   Connection &signal&idler &User 1&User 2  \\ 
                \hline
     top &  Alice-Satellite &$\lambda_1$&$\Lambda_1$& - &-\\
      
     top &   Bob-Satellite&$\lambda_2$&$\Lambda_2$& -&- \\
     top &  Charlie-Satellite &$\lambda_3$&$\Lambda_3$& - &-\\
     top  &  Dana-Satellite&$\lambda_4$&$\Lambda_4$ & -&-\\
      middle  &   Alice-Charlie &-&-&$\Lambda_1,\lambda_3 $ & $\lambda_1,\Lambda_3$  \\
           middle   &  Bob-Charlie &-&-&$\Lambda_2$  & $\lambda_2$ \\
     bottom  & Alice-Bob&-&- &$\Lambda_1,
    \lambda_2$   & $\lambda_1,\Lambda_2$  \\
    bottom   &  Bob-Charlie &-&-&$\lambda_3$  & $\Lambda_3$ \\

                \hline
    \end{tabular}

    \label{tab:alloc}
\end{table}

\begin{figure}
\begin{subfigure}{0.49\textwidth}
\captionsetup{font=bf}
    \centering
    \includegraphics[width=\textwidth]{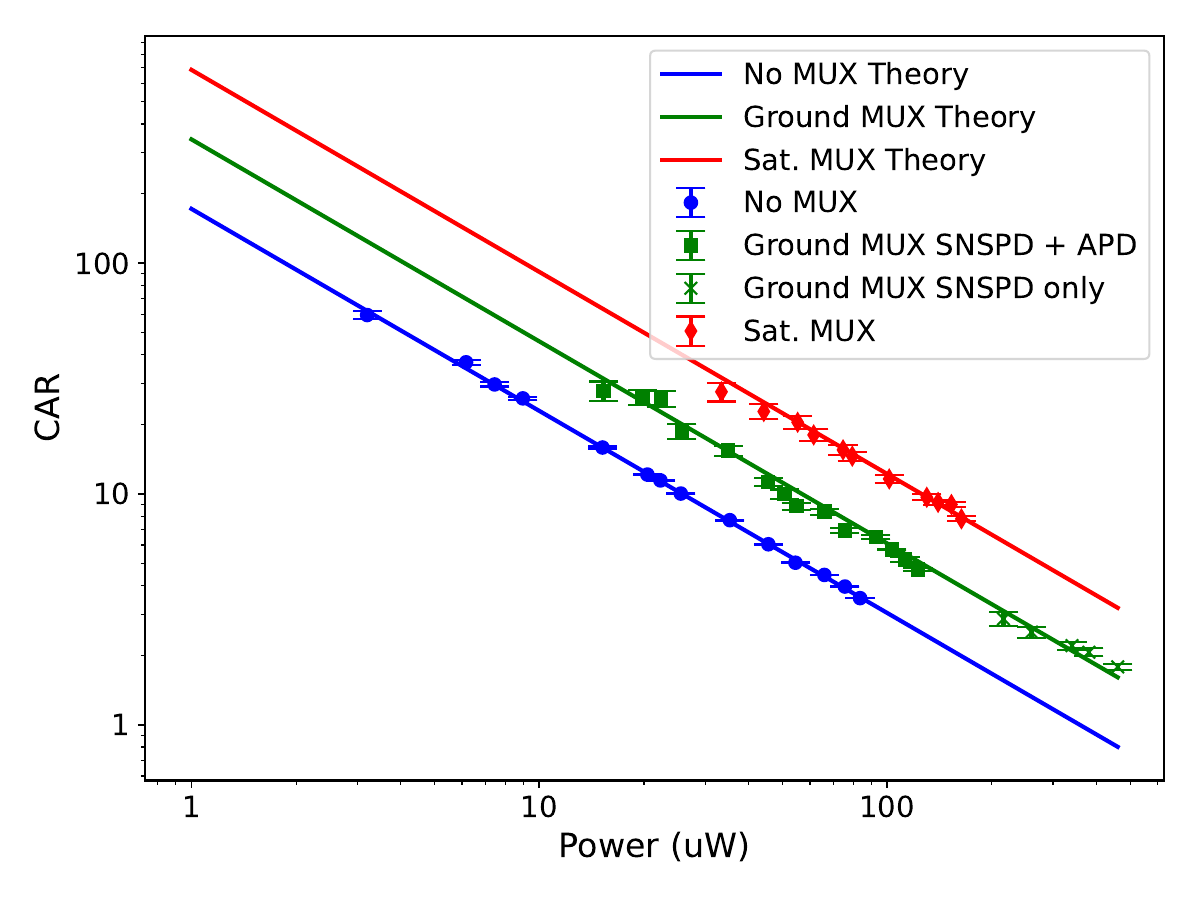}
    \caption{}
    \label{fig:CAR}
    \end{subfigure}
    \begin{subfigure}{0.49\textwidth}
    \captionsetup{font=bf}
    \includegraphics[width=\textwidth]{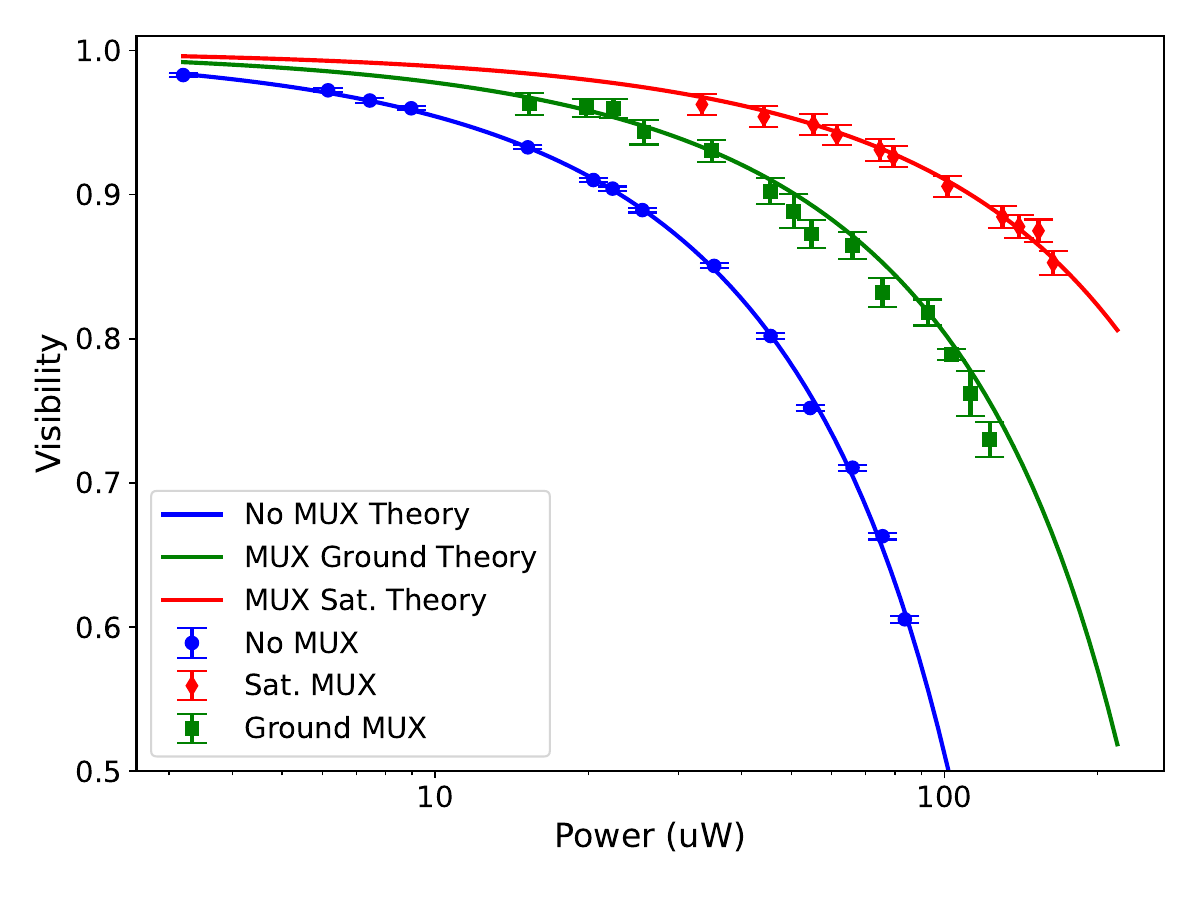}
    \caption{}
        \label{fig:Vis}
\end{subfigure}
\begin{subfigure}{0.49\textwidth}
\captionsetup{font=bf}
    \centering
    \includegraphics[width=\textwidth]{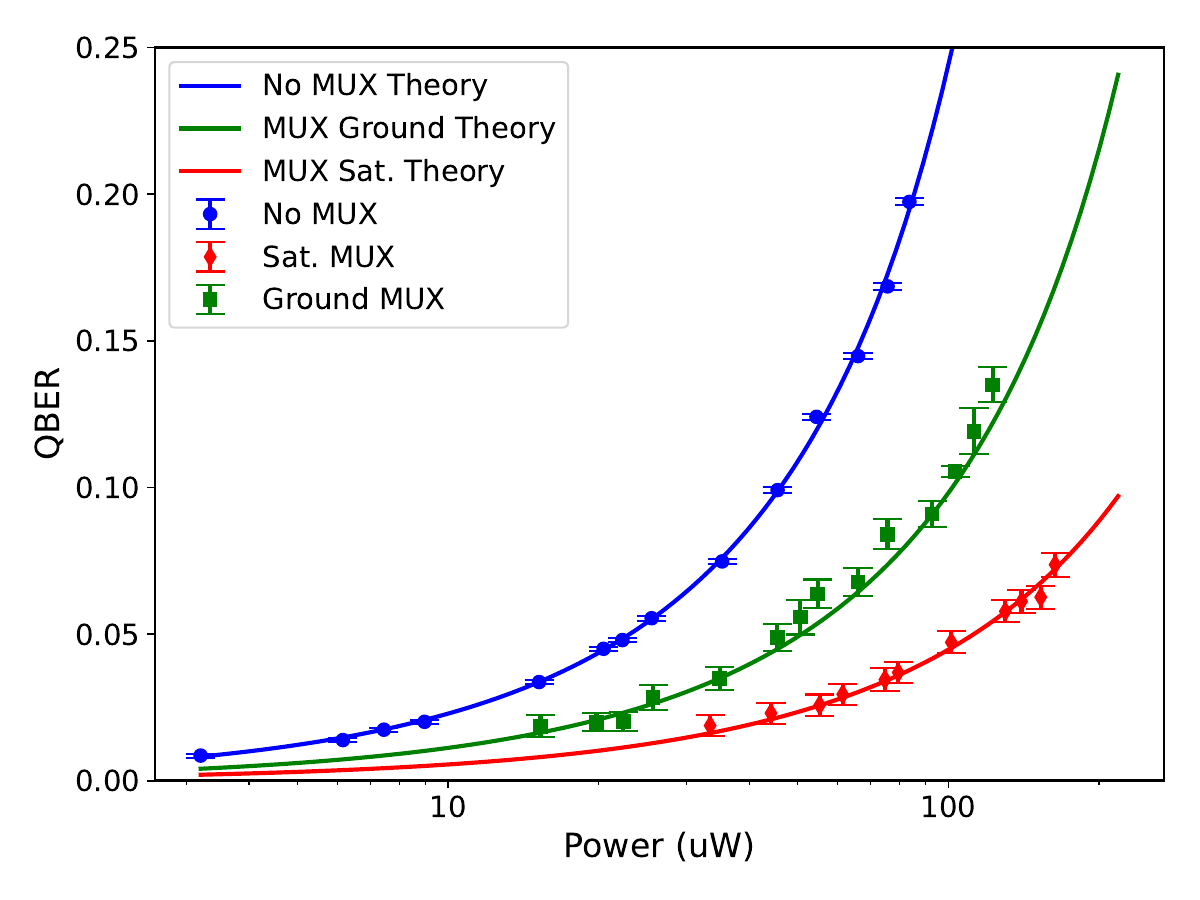}
    \caption{}
    \label{fig:QBER}
    \end{subfigure} 
    \begin{subfigure}{0.49\textwidth}
    \captionsetup{font=bf}
    \includegraphics[width=\textwidth]{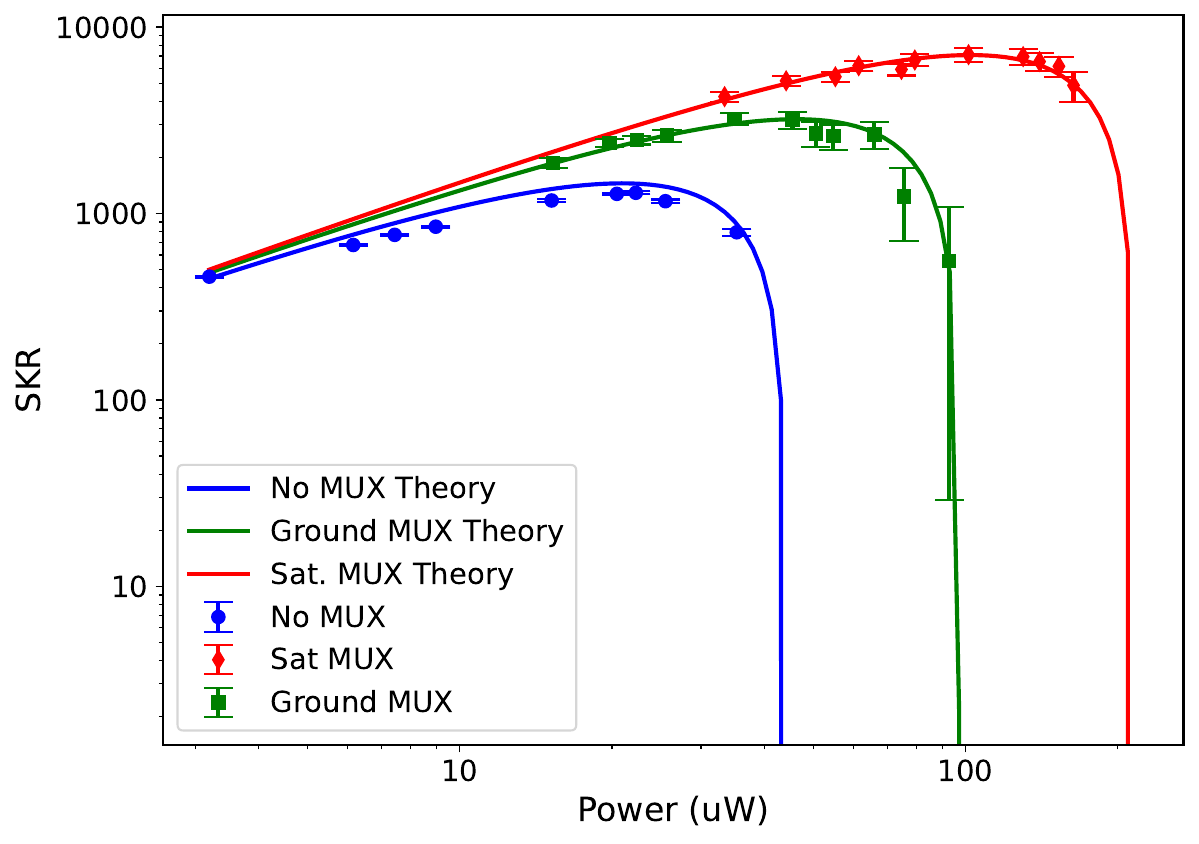}
    \caption{}
        \label{fig:SKR}
\end{subfigure}
\caption{Multiplexing scheme performance estimation. In \textbf{(a)}, the coincidence-to-accidentals ratio is measured as a function of the pump power. In \textbf{(b)}, \textbf{(c)}, and \textbf{(d)} the two-photon interference visibility, quantum bit error rate (QBER), and secure key rate are respectively estimated from the measured CAR. For the ground scenario shown in \textbf{(a)}, coincidences at low pump power were measured using a SNSPD and APD detector per user, due to the SNSPD's poor detection efficiency at 780 nm. At higher pump power, signal and idler photons were jointly detected on a single SNSPD channel per user. Here, \textit{MUX} refers to the multiplexed configuration of the network, while \textit{No MUX} denotes the single-channel baseline.} 
\end{figure}

The CAR was calculated from $\rm{CAR}=R_m/R_{um}$, where $R_m$ and $R_{um}$ are the coincidence rates at matched and un-matched slots as defined in Ref.~\cite{Takesue_2005}. The former corresponding to true coincidences caused by photons generated in the same pump pulse whereas the latter represents accidental coincidences arising from multi-pair emissions or detector dark counts. Exploiting time and frequency multiplexing at the satellite and ground nodes, respectively, accidental coincidences, due to uncorrelated channel detections, are filtered, leading to a CAR improvement in accordance with the theoretical expectation as shown in Fig.~\ref{fig:CAR}. More explicitly, one can calculate the CAR from the correlated photon pair rate ($\mu_c$), the transmittance of the signal ($\alpha_s)$ and idler channels ($\alpha_i$) and the average signal $(c_s)$ and idler $(c_i)$ count rate according to \cite{Takesue_2005}:
\begin{equation}
    \rm{CAR}=\frac{\mu_c\alpha_s\alpha_i}{c_sc_i}+1 .
\end{equation}
Assuming $N$ multiplexed channels, $\rm{CAR}-1\xRightarrow{MUX}\frac{(\mu_c/N)\alpha_s\alpha_i}{(c_s/N)(c_i/N)} = \frac{N\mu_c\alpha_s\alpha_i}{c_sc_i}$ resulting in a linear CAR improvement \cite{Lohrmann_2020}. For the satellite MUX, the reduced chi-squared relative to the theoretical curve is $\chi^2_{\rm{sat}} = 1.14$, indicating excellent agreement within the reported uncertainties. For the ground MUX, the best-fit scaling is 1.91$\times$ slightly below the theoretical expectation, giving $\chi^2_{\rm{ground}} = 2.8$. This small deviation can be attributed to slight cross-talk between the multiplexed channels and limited coincidence statistics resulting from significant losses in the demultiplexing stage. To limit the cross-talk between the channels, 1 nm guard bands between the signal channels were used to limit their spectral overlap. 
\par From the CAR, we approximate the two-photon interference visibility using the relation $V=\frac{\rm{CAR}-2}{\rm{CAR}-1}$ derived in \cite{TAKESUE2010276}. The multiplexing scheme enhances the visibility in Fig.~\ref{fig:Vis}, allowing increased coincidence rates at higher pump powers. In our analysis, the inferred visibility accounts only for accidental coincidences arising from multi-pair emissions and background counts, with no additional error channels assumed. To contextualize these results for practical applications, we estimate the quantum bit error rate (QBER) according to $\rm{QBER}=\frac{1-V}{2}$ and the asymptotic secure key rate (SKR) for the entanglement-based quantum key distribution (QKD) protocol outlined in \cite{Ma, Holloway_paper}, \begin{equation}\label{eq:SKR}
    SKR = q\{Q_\lambda[1-(1+f_e)H_2(\rm{QBER})]\}
\end{equation}
where $q=1/2$ is the basis reconciliation factor for BBM92, $Q_\lambda$ is the overall gain estimated from the coincidence counts, $f_e$ is the bidirection error correction efficiency (here taken as $f_e=1.11$) and $H_2(x)$ is the binary entropy function,
\begin{equation}
    H_2(x)=-x\log_2(x)-(1-x)\log_2(1-x).
\end{equation}
These results are respectively presented in Figs.~\ref{fig:QBER}-\ref{fig:SKR}. Again, multiplexing allows for a reduction of the QBER and a higher attainable secure key rate. This method thus presents a scalable approach to enhance the entanglement generation rate in a multipoint-to-point quantum link during the satellite pass. Moreover, additional users can be incorporated to the ground network without compromising the quality of service or pairwise entanglement generation rate - and with minimal resource overhead. By allocating a distinct temporal window to each channel within the pulse train \cite{wengerowskyentanglementbased2018}, each user can unambiguously discriminate between frequency channels using a single detector. Accordingly, coincidences, for the ground configuration at high pump power in Fig.~\ref{fig:CAR}, were measured with only one detector channel per user. Despite the limited detection efficiency of the SNSPD at 780 nm ($\eta_{780}\sim6\%$), temporally resolving each channel still yields a twofold CAR improvement.  

 \section{Discussion}\label{discussion}
We have experimentally demonstrated a reconfigurable quantum network architecture first proposed in 
Ref.~\cite{vinet_reconfigurable_2025}. In contrast to previous demonstrations, which are limited to metropolitan distances \cite{Joshi, wengerowskyentanglementbased2018, liu40user2022, qi15user2021,krzic_towards_2023,chen_metropolitan_2010}, our architecture is specifically designed to integrate satellite links, enabling a hybrid terrestrial–satellite quantum network \cite{chenintegrated2021}. Leveraging both time and frequency multiplexing, we demonstrate an increased coincidence-to-accidental ratio without additional resource overhead. Our approach is conceptually similar to that proposed by Wengerowsky \textit{et al} \cite{wengerowskyentanglementbased2018}; however, their approach requires an external gating signal, making it unsuitable for mobile nodes. Here, using frequency-to-time mapping, we circumvent this limitation, enabling fully passive detection compatible with dynamic satellite passes. Moreover, as all channels are defined within the same entangled source and each user requires a single detector channel, the network scales linearly.
\subsection*{Scaling potential}
\begin{figure}
    \centering
    \includegraphics[width=0.6\linewidth]{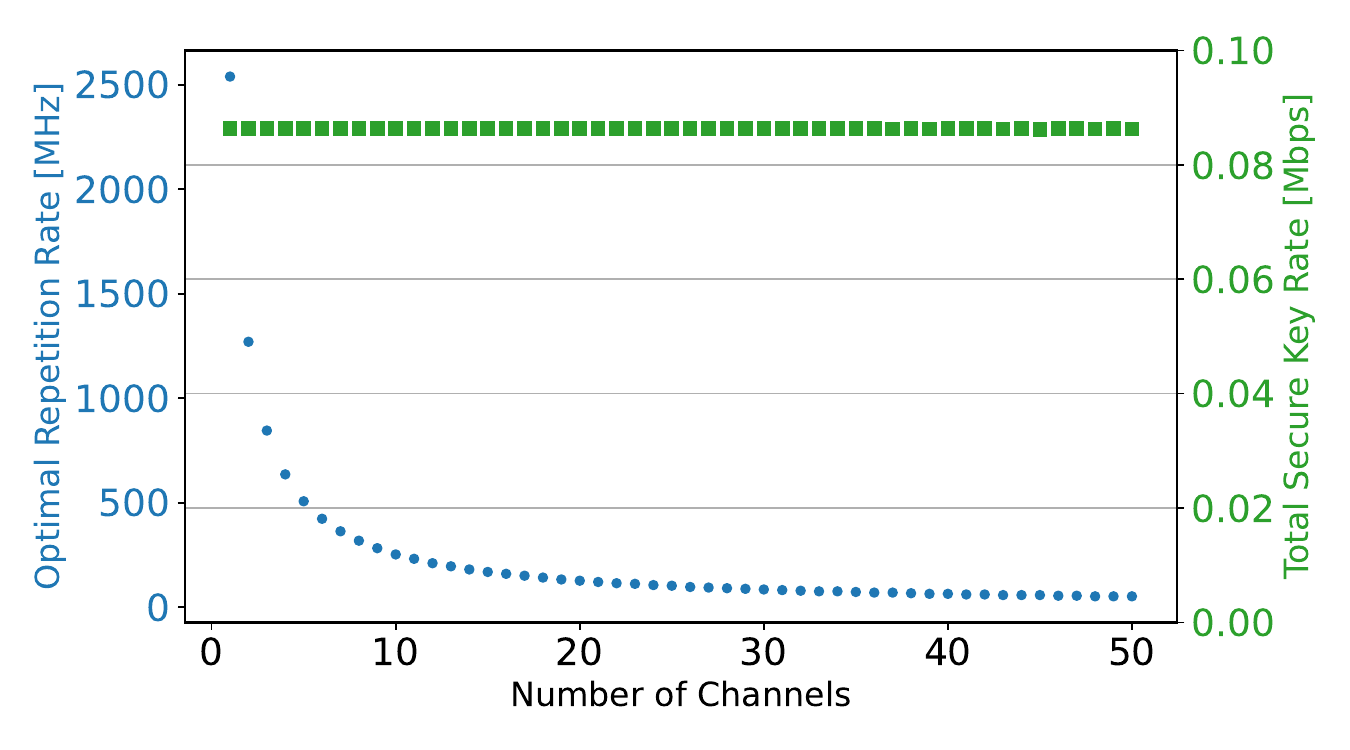}
    \caption{Trade-off between the number of frequency-time multiplexed channels and the repetition rate assuming a pump power of $100$ $\mu$W, 30 dB channel loss, detector dead time of 1 $\mu$s, a dark-count rate of 1000 counts/s, and a timing jitter of $130$ ps.}
    \label{fig:trade}
\end{figure}

Photon bandwidth and detector timing resolution underpin, respectively, the capacity for wavelength and time multiplexing. Consequently, for the satellite configuration, the number of channels is bounded by $(R\times\delta t_j)^{-1}$, where $R$ is the pulse repetition rate and $\delta t_j$ is the timing resolution of the detection system, including synchronization-induced jitter. There is a clear compromise between the pulsed laser repetition rate and the number of available time-multiplexed channels. This trade-off is illustrated in Fig.~\ref{fig:trade} for our source, assuming a pump power of 100 $\mu$W, channel loss of 30 dB, detector dead time of 1 $\mu$s, 1000 counts per second dark count rate and timing jitter of 130 ps.
Owing to the requirement for spectral compatibility with the QEYSSat payload, the signal bandwidth of the ground configuration is restricted to the satellite acceptance window of 780–795 nm. Nevertheless, in contrast to the planned primary QEYSSat source \cite{PhysRevA.110.063515}, which is limited by its intrinsically narrow spectral bandwidth and supports essentially single-channel operation, our configuration leverages the full 15 nm acceptance window. Assuming 100 GHz spacing between channels, this  could be split between 72 parallel channels. In low-loss conditions, this could allow up to 72 separate users to access the same satellite link. However, in the presence of loss and noise, the key rate for each user may become impractically low. In this case, dynamic reconfiguration of the idler demultiplexing would send multiple channels to each user ensuring that a smaller number of users can access the link, but with higher key rates.

\begin{figure}
    \centering
    \includegraphics[width=\linewidth]{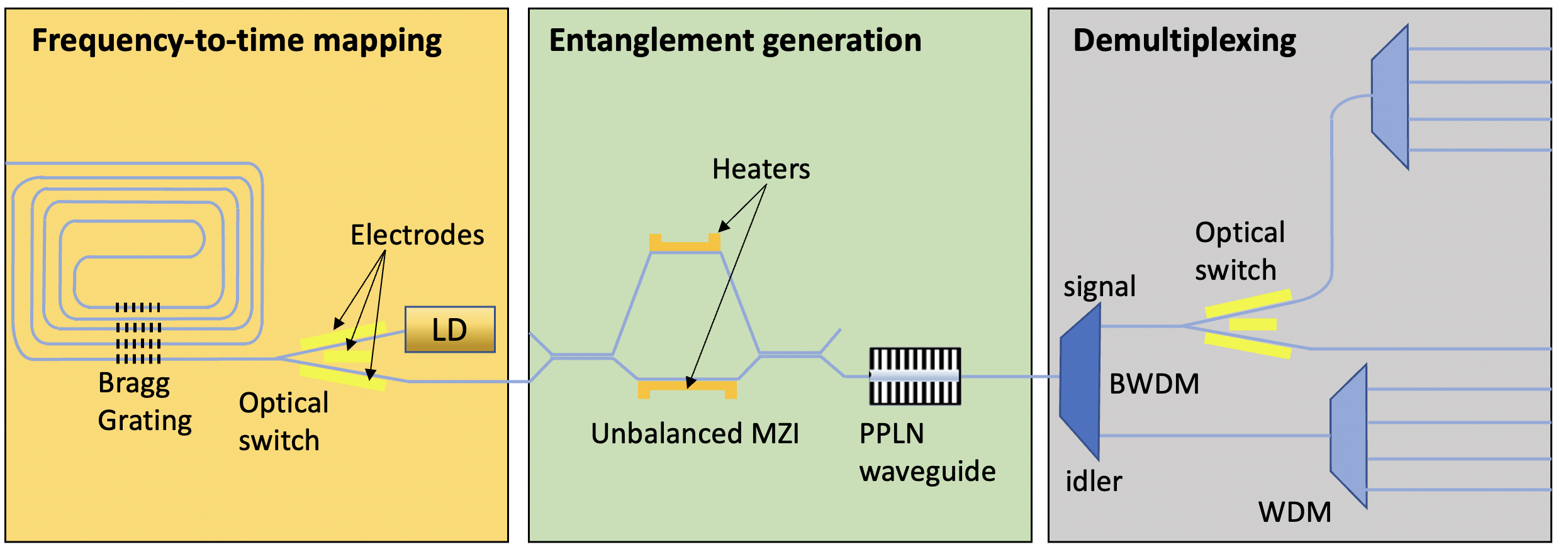}
    \caption{Schematic for the proposed integrated networkable quantum source. }
    \label{fig:integrated}
\end{figure}
\subsection*{Considerations for practical deployment}
The network could be further optimized by using flexible grid technology \cite{Lingaraju_2021}: employing a wavelength selective switch or reconfigurable optical add-drop multiplexer \cite{Wang_2017} would not only simplify the experimental design, but also enable adaptive bandwidth management beyond the two configurations showcased in Fig.~\ref{fig:setup}. Moreover, under flex-grid channel allocation, entanglement distribution losses are fixed and independent of the number of users. This is in contrast to our implementation, where FBG-circulator stages are cascaded, leading to an incremental loss per added user due to the additional components required (see Appendix~\ref{sec:Materials}). Flex-grid operation thus provides a significant scalability advantage over the standard WDM approach. 
\par Evidently, a field-deployed demonstration would encounter varying link conditions with significantly higher losses, on the order of ~40 dB for satellite up-links with varying loss throughout the satellite pass \cite{Bourgoin_2013}. Nevertheless, the CAR scaling is resilient to loss, as additional attenuation reduces both the coincidence and singles rates proportionally, leaving the ratio unchanged until the background-noise floor is reached. For a 25 cm aperture receiver telescope, such as on the QEYSSat payload, recent up-link estimates report background levels on the order of kilocounts per second under nighttime urban conditions \cite{yastremski_estimating_2025}. In this high loss regime, the impact of detector dead time is minimal because the detected count rates remain well below the saturation threshold \cite{vinet_reconfigurable_2025,PhysRevA.106.062607}. Conversely, in the low-loss fiber link, detector dead time can be mitigated by distributing the photon flux across multiple detection channels thereby suppressing dead-time-induced saturation effects. Atmospheric turbulence induces phase noise, perturbing both temporally and spatially the optical wavefront \cite{PhysRevLett.133.020201}. The resulting fluctuations in the arrival time of the photons add to the timing jitter of the detection system, potentially limiting the resolvable number of frequency-time multiplexed channels. In practice, however, the contribution is typically minimal for satellite up-links, on the order of 0.5 mm RMS \cite{Kral:05}.
Moreover, the reconfigurable architecture is well-suited for dynamic and challenging environmental conditions. When the free-space link becomes unavailable, due to poor weather or significant cloud coverage, the system can be switched to the ground-based configuration thereby ensuring continued network uptime. 

\subsection*{Integrated design}
  \par
\begin{figure}
      \centering
      \includegraphics[width=0.5\linewidth]{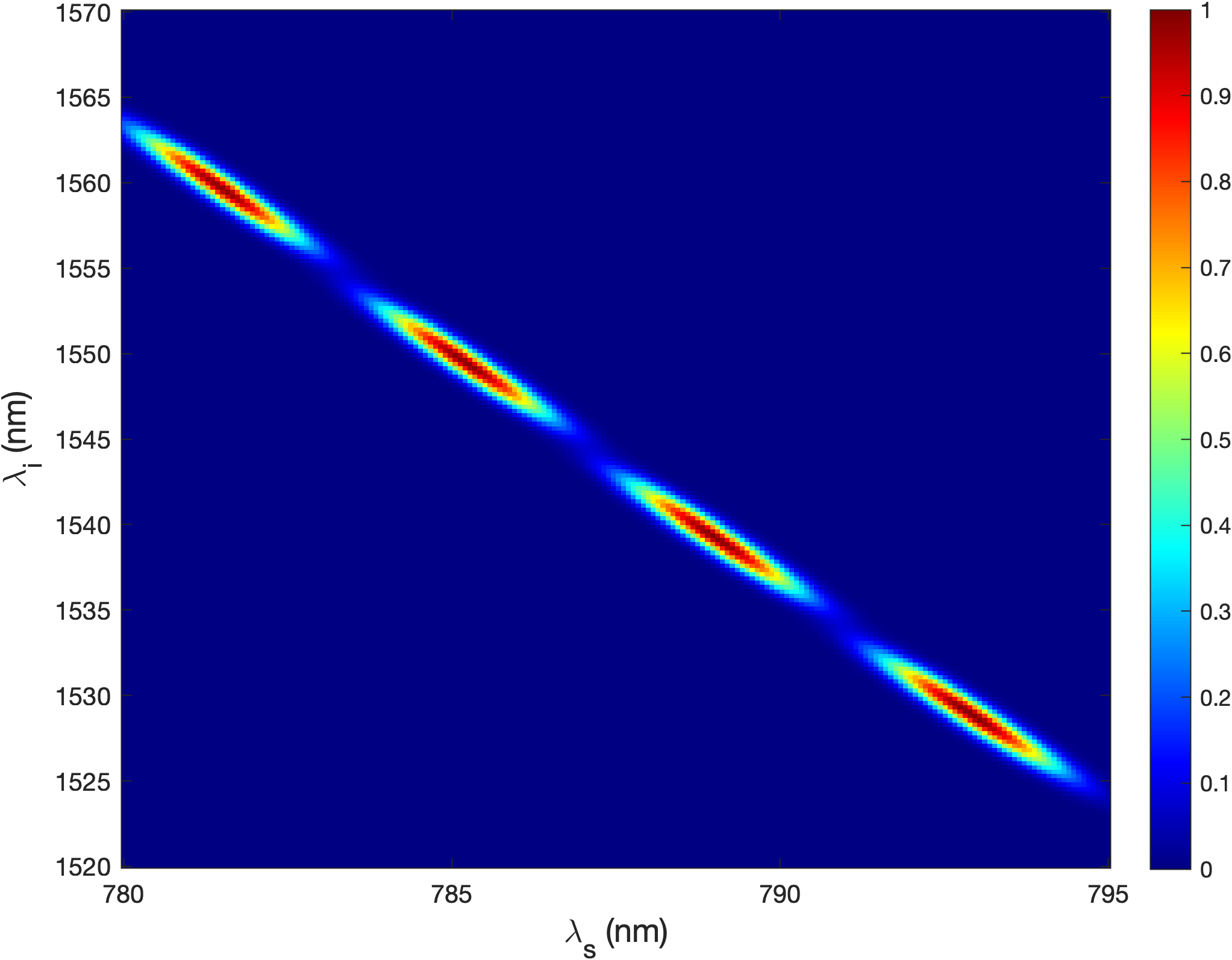}
      \caption{Simulated joint spectrum intensity (JSI) for the entangled photon source depicted in Fig.~\ref{fig:integrated}. Each JSI represents the spectral correlations between signal and idler photons for a specific frequency-time channel.}
      \label{fig:JSI}
  \end{figure}

 Compact, adaptable, and cost-effective technologies will be essential to scale quantum networks toward global deployment. Recent work on integrated photonic entangled photon sources has shown significant improvements in stability, energy efficiency, and miniaturization paving the way toward scalable quantum networking systems \cite{appas_flexible_2021,zheng_multichip_2023,PhysRevX.8.021009}. Motivated by these efforts, we propose in Fig.~\ref{fig:integrated} an integrated entangled photon source design. The design emphasizes practicality and compatibility with satellite-enabled quantum networks. To minimize size and cost, the titanium-sapphire in Fig.~\ref{fig:setup} is replaced by a gain-switched laser diode that produces broadband picosecond optical pulses. While femtosecond pulses could be used with time-bin encoding, care must be taken to manage chromatic dispersion and group-velocity mismatch in the nonlinear crystal and any fibers. The use of picosecond pulses in our integrated design provides a practical balance: reduced sensitivity to dispersion while maintaining sufficient bandwidth for broadband SPDC and wavelength multiplexing.
    The FTM is applied on the pump field using a Y-branched digital optical switch \cite{Singh2010DOS} in combination with a spiral waveguide containing an array of on-chip Bragg gratings. 
     By encoding the FTM on the pump photons, the insertion loss associated to its optical components can be compensated. After reflection, the temporally stretched pulse is directed through a thermally tuned unbalanced Mach-Zehnder interferometer, where a coherent superposition of early and late time-bins is encoded. The resulting pulse is then coupled into a PPLN waveguide, where time-bin entangled signal and idler photon pairs are generated via SPDC. These photon pairs are subsequently separated by a band-wavelength division multiplexer (BWDM). A second optical switch toggles between the satellite and ground configurations, rerouting the signal to a WDM for the latter. To evaluate the required guard bands for suppressing cross-talk in the FTM stage, Fig.~\ref{fig:JSI} shows a simulated joint spectral intensity calculated from the SPDC process of the proposed integrated entangled photon source. Using parameters consistent with the demonstration presented in section~\ref{sec:exp}, a $1.75$ nm pump was divided into four frequency channels, each $0.25$ nm wide. $0.25$ nm spectral gaps were inserted between the Bragg grating reflection bands. This configuration was found to reduce interchannel cross-talk to below $-30$ dB, ensuring that the down-converted photon pairs generated from different pump channels remain spectrally and temporally well separated. 
This design could easily be extended to additional channels by increasing the number of Bragg gratings in the FTM unit. Such a source would enable quantum communication nodes that are portable, power-efficient, and suitable for deployment in real-world settings. The design consolidates the FTM \cite{wang_reconfigurable_2015,Singh:23}, phase control, nonlinear photon-pair generation \cite{guo_parametric_2017,PhysRevLett.113.103601} and demultiplexing onto a single photonic chip \cite{li_polarization-independent_2023,piggott_inverse_2015}, significantly reducing size, alignment overhead, and sensitivity to environmental noise. 


\begin{backmatter}
\bmsection{Funding}
This work is supported by the  High Throughput and Secure Networks (HTSN) challenge by the National Research Council Canada, project HTSN-630. The authors further acknowledge support from the Natural Sciences and Engineering Research Council of Canada (NSERC), the Canadian Foundation for Innovation (CFI), the Ontario Research Fund (ORF) and the Canada Excellence Research Chair program (CERC). SV would like to thank the NSERC CGS-D for personal funding.
\bmsection{Acknowledgments}
The authors acknowledge helpful discussions with Ramy Tannous, Julien Beaudoin Bertrand, and Mamoun Wahbeh. 
\bmsection{Disclosures} 
The authors declare no conflicts of interest related to this article.
\end{backmatter}
\appendix
\section{Materials and Methods}\label{sec:Materials}
\subsection*{Waveguide coupling}
 To enable precise positional adjustment, the input (\textit{OZ Optics} 532 nm PM fibre) and output (1550 nm PM lensed fibre) fibres were mounted on on three-axis translation stages (\textit{MAX313D, Thorlabs}), each controlled by a piezo controller (\textit{BPC203, Thorlabs}).
The input fibre was flat cleaved and butt coupled to the waveguide which was mounted in its oven (\textit{HC Photonics}) and placed on a two-axis (vertical and horizontal) translation stage for precise positioning. The coupling process was monitored via an overhead stereo microscope (\textit{Leica M125 C}) with a tunable magnification between 12.8 and 160X \cite{Vinet2025_SatelliteAssistedQuantumComm}. Coupling into and out of the waveguide introduced 4.5 dB loss. The waveguide used in the experiment was a PPLN ridge waveguide fabricated on a z-cut substrate. The chip dimensions were $15\times0.5\times1.5$ mm (length $\times$ width $\times$ thickness), with the waveguide defined on the z-facet at a depth of approximately $2.8 \mu$m. The width of the channel was $6 \mu$m and the mode field diameter were $4.83$ and $4.16\mu$m for the transverse y- and z-directions respectively. Following the down conversion, a free-space bridge separated the signal and idler photons, with measured coupling efficiencies of 44$\%$ (3.57 dB loss) for idler photons and 16$\%$ (7.96 dB loss) for signal photons. The reduced coupling efficiency for the latter was due to spatial mode filtering from the PM780 fibre (\textit{Thorlabs}) as the $\sim780$ nm light is slightly multi-modal in the $1550$ nm fibre at the output of the waveguide. 
\subsection*{Frequency-to-time mapping}
The FTM unit consists of an array of four fibre Bragg gratings, each designed to reflect one spectral channel $\lambda_1-\lambda_4$ as indicated in Fig.~\ref{fig:spectrum}a. More precisely, the four FBGs (\textit{O/E Land Inc.}) have central wavelengths at $781.5, 785.5, 789.5$ and $793.5$ nm, each with a bandwidth of $3$ nm, and separated by $35$ cm of PM-780-HP fibre. As each grating is positioned at a different location along the fibre, the reflected photons traverse an additional round-trip optical path before returning, resulting in a wavelength-dependent time delay of approximately $3.4$ ns between adjacent channels. By mapping each frequency channel to a distinct arrival time, the FTM unit enables the satellite node to resolve multiple channels with a single detector, reducing hardware complexity while maintaining full channel discrimination. The FTM unit added 10 dB loss (3 dB from the optical circulator and 7 dB from the FBG array). For the satellite configuration, the FTM unit and the Si-APD are connected via a 5 meter 780 PM fiber (\textit{Thorlabs 780PM}). Taking into account the detection efficiency, the total loss of the signal channel for the satellite configuration is $\sim26$ dB. 
\subsection*{Wavelength division demultiplexing}
The ground users are separated by 10 meters of telecom optical fiber (\textit{Thorlabs SMF-28}). Demultiplexing for the idler photons was implemented via a cascaded assembly of circulators and FBGs, resulting in losses of 4.9 dB, 7.7 dB, 10.5 dB, and 8.5 dB for Alice, Bob, Charlie, and Dana, respectively. Adding the 8 dB contribution from the waveguide loss and free-space bridge, the total idler-arm losses for the satellite configuration amount to approximately 12.9 dB, 15.7 dB, 18.5 dB, and 16.5 dB for Alice, Bob, Charlie, and Dana, respectively. For the ground configuration, the 1550/780 WDM had an insertion loss of 0.97 and 3.02 dB at 1550 and 780 nm respectively. The individual signal fiber Bragg gratings ($\rm{FBG_{s,2},FBG_{s,3}}$) had a loss of $\sim 1.5$ dB. The Si-APD had an efficiency $>60\%$ at 780\,nm. The SNSPD had an efficiency of $>75\%$ at 1550\,nm but just $\sim6\%$ at 780\,nm. However SNSPDs that are efficient at both wavelengths are already under development. \cite{Gourgues:19,doi:10.1021/acsphotonics.0c00433}.

\bibliography{sample}

\end{document}